\documentclass[onecolumn, draft, 12pt]{IEEEtran}
\usepackage{amsmath,cases}
\usepackage{amsthm}
\usepackage{amssymb}
\usepackage{cite}
\usepackage{amsfonts}
\usepackage{enumerate}
\usepackage[final]{graphicx}
\usepackage{multirow}
\allowdisplaybreaks

\newcommand{\vrho}{\ensuremath{\varrho}}

\newcommand{\E}{{\rm E}}
\newcommand{\Cs}{\mathbb{B}}
\newcommand{\Ni}{\mathbb{N}_{i}}
\newcommand{\Ns}{\mathbb{N}}
\newcommand{\Bs}{\Cs}

\newcommand{\Cprime}{\Cs^{'}}
\newcommand{\Ss}{\Cs}
\newcommand{\G}{\mathbb{G}}
\newcommand{\R}{\mathbb{R}}

\newcommand{\Rn}{\mathbb{R}^{N}}
\newcommand{\Rnm}{\mathbb{R}^{(N-1)}}
\newcommand{\Rmn}{\mathbb{R}^{N \times (N-1) }}
\newcommand{\Rnmm}{\mathbb{R}^{(N-1) \times (N-1)}}

\newcommand{\mse}{\xi_{_{N}}}

\newcommand{\nld}{\rm NLC}
\newcommand{\rnld}{\rm RC}

\newcommand{\Lg}{\mathcal{L}}

\newcommand{\Ltwo}{\mathcal{L}_{2}}
\newcommand{\dmp}{\mathcal{X}}

\newcommand{\onevect}{\mathbf{1}}
\newcommand{\onevectT}{\mathbf{1}^{\mathrm{T}}}

\newcommand{\La}{\mathbf{L}}
\newcommand{\D}{\mathbf{D}}
\newcommand{\B}{\mathbf{B}}
\newcommand{\A}{\mathbf{A}}
\newcommand{\I}{\mathbf{I}}

\newcommand{\Pm}{\mathbf{M}}

\newcommand{\e}{\mathbf{e}}

\newcommand{\Xcp}{\mathbf{x}_{\Cs \perp}}

\newcommand{\mXc}{\mathbf{\boldsymbol\mu}_{\Cs}(\mathbf{x})}
\newcommand{\mXcp}{\mathbf{\boldsymbol\mu}_{\Cs \perp}(\mathbf{x})}

\newcommand{\U}{\mathbf{U}}
\newcommand{\Sgma}{\mathbf{\Sigma}}

\newcommand{\UT}{\mathbf{U}^{\mathrm{T}}}
\newcommand{\XPX}{{\mathbf{x}}^{\mathrm{T}} \mathbf{M} \mathbf{x}}
\newcommand{\XTMXt}{{\mathbf{X}(t+1)}^{\mathrm{T}} \mathbf{M} \mathbf{X}(t+1)}

\newcommand{\Czero}{\mathbf{C}}

\newcommand{\CSR}{\Czero_{\rm RC}}

\newcommand{\So}{\mathbf{S}}

\newcommand{\Stheta}{\So^{\theta_0}}
\newcommand{\fiN}{\mathbf{\Phi}}
\newcommand{\fiNT}{\mathbf{\Phi}^{\mathrm{T}}}

\newcommand{\xbar}{\bar{x}}

\newcommand{\x}{\mathbf{x}}
\newcommand{\X}{\mathbf{X}}
\newcommand{\xX}{\mathbf{x}}

\newcommand{\Z}{\mathbf{z}}

\newcommand{\cval}{\ensuremath{\theta^{*}}}

\newcommand{\mx}{\mathbf{\boldsymbol\mu}(\mathbf{x})}

\newcommand{\mXt}{\mathbf{\boldsymbol\mu}(\mathbf{X}(t))}

\newcommand{\nv}{\mathbf{n}}

\newcommand{\vX}{ V(\mathbf{x})}

\newcommand{\vecXtone}{ V(\mathbf{X}(t+1))}
\newcommand{\fitX}{ \varphi(\mathbf{x})}

\newcommand{\ntilde}{\tilde{\mathbf{n}}}
\newcommand{\ntildet}{\tilde{\mathbf{n}}(t)}
\newcommand{\xtilde}{\tilde{x}}
\newcommand{\xtildet}{\tilde{x}(t)}
\newcommand{\Xtilde}{\tilde{\mathbf{X}}}
\newcommand{\Xtildet}{\tilde{\mathbf{X}}(t)}

\newcommand{\itN}{1 \leq i \leq N}

\newtheorem{thm}{Theorem}

\newtheorem{lem}{Lemma}

\begin{document}
\title{Robust Consensus in the Presence of Impulsive Channel Noise}
\author{Sivaraman Dasarathan, Cihan Tepedelenlio\u{g}lu, \emph{Member, IEEE}, Mahesh Banavar, \emph{Member, IEEE} and Andreas Spanias, \emph{Fellow, IEEE}
\thanks{The authors are with the School of Electrical, Computer, and Energy Engineering, Arizona State University, Tempe, AZ 85287, USA. (Email: \{sdasarat, cihan, mbanavar, spanias\}@asu.edu). This work was supported in part by the National Science Foundation under Grants NSF FRP 1231034 and NSF CCSS 1307982.}
} \maketitle
 
\begin{abstract}
A distributed average consensus algorithm robust to a wide range of impulsive channel noise distributions is proposed. This work is the first of its kind in the literature to propose a consensus algorithm which relaxes the requirement of finite moments on the communication noise. It is shown that the nodes reach consensus asymptotically to a finite random variable whose expectation is the desired sample average of the initial observations with a variance that depends on the step size of the algorithm and the receiver nonlinear function. The asymptotic performance is characterized by deriving the asymptotic covariance matrix using results from stochastic approximation theory. Simulations corroborate our analytical findings and highlight the robustness of the proposed algorithm.
\end{abstract}

\section{Introduction} \label{sec:intro_btx_consensus_robust}
Wireless sensor networks (WSNs) without a fusion center have the advantages of robustness to node failures and being able to function autonomously without a central node controlling the entire network \cite{Sankarasubramaniam2002}. In such fully distributed networks, sensors collaborate with their neighbours by repeatedly exchanging information which they combine locally to achieve a desired global objective. For example, the sensors could come to an agreement on the sample average (or on a global function) of initial measurements. This is called distributed consensus. Distributed consensus algorithms have attracted significant interest in the recent past and have found several applications in areas such as healthcare, environmental monitoring, military and home appliances \cite{Boyd2003,Boyd2004,OlfatiSaber2003,OlfatiSaber2007,MinyiHuang2008,Oreshkin2008,KarMoura2009}. 

In existing literature on consensus in the presence of communication noise, the additive noise is always assumed to have finite moments \cite{Boyd2007,Touri2009,MinyiHuang2007,Pescosolido2008,Barbarossa2008,MinyiHuang2008,AysalBarner2010,Nedic2011cvx, KarMoura2009,KarMoura2007}. Sensor networks which operate in adverse conditions can be susceptible to impulsive noise distributions. For example, the aggregated interference at a desired node from its neighbouring nodes of a Poisson network is characterized by alpha-stable distribution which may not have finite mean or variance \cite{Sousa1992,Ilow1998,Yang2003,Hughes2000,Haenggi2009,Win2009,JunghoonLee2011,RajanTep2010}. Therefore there is a need to develop consensus algorithms which are robust to impulsive channel noise. Consensus with nonlinear combining at the receiver has been considered in \cite{KhanKar,OlfatiSaber2003,Ulrich2008,HuiWassim2008,WenwuChen2011,Ajorlou2011} only in the absence of inter-sensor communication noise. Therefore, it is of interest to solve the problem of distributed consensus with receiver nonlinearities that soft-limit the impulsive additive noise.

In this paper, we propose a robust consensus ($\rnld$) algorithm which is robust to impulsive communication noise by soft-limiting at receiver sensor nodes before combining. We do not require the channel noise to have finite moments as is assumed in all the previous work on distributed average consensus algorithms \cite{Boyd2007,Touri2009,MinyiHuang2007,Pescosolido2008,Barbarossa2008,MinyiHuang2008,AysalBarner2010,Nedic2011cvx, KarMoura2009,KarMoura2007}. In addition, like in \cite{dastep2013}, we assume that every sensor maps its state value through a bounded function before transmission to respect a peak power constraint at every iteration making it ideal for resource-constrained WSNs. We prove that all the sensors employing the $\rnld$ algorithm reach consensus to a finite random variable whose mean is the desired sample average. We characterize the asymptotic performance by deriving the asymptotic covariance matrix using results from stochastic approximation theory. Finally, we explore the performance of the proposed algorithm employing various functions for the transmit and receiver non-linearities. Different from \cite{MinyiHuang2008,KarMoura2009} and \cite{KarMoura2007} which also considered consensus in the presence of noisy transmissions, herein we analyse nonlinear processing both at the transmit and receiver nodes and study the asymptotic covariance matrix and its dependence on both the power-constraining transmit nonlinearity, and the soft-limiting receive nonlinearity. It is shown that the norm of the asymptotic covariance matrix is limited by the Fisher information of the noise distribution with respect to a location parameter.

The rest of this paper is organized as follows. We begin by reviewing network graph theory in Section \ref{sec:review_spectral_robust}. In Section \ref{sec:consensus_no_noise_robust}, we describe the sensing and channel models and introduce the consensus problem. We consider the $\rnld$ algorithm in the presence of noise in Section \ref{sec:consensus_with_noise_robust}, and prove that the sensors reach consensus to a random variable. In Section  \ref{sec:simulations_nld_robust}, we present several simulation examples to study the performance of the proposed algorithm. Concluding remarks are presented in Section \ref{Sec:Conclusions:consensus_robust}. 

\subsection*{Notations and Conventions}\label{subsec:nld_notations_robust}
Vectors are denoted by boldface upper-case or lower-case letters and matrices are denoted by boldface upper-case letters. $\max \lbrace a_1 , a_2\rbrace$ denotes the maximum of $a_1$ and $a_2$. ${\rm diag} [a_{1},\;  a_{2}, \; \ldots, \; a_{N}]$ denotes an $N \times N$ diagonal matrix whose diagonal elements are given by $a_{1}, a_{2}, \ldots, a_{N}$. $\E[\cdot]$ denotes the expectation operator. The symbol $\| \cdot \|$ denotes the ${l}_{_{2}}$ norm for vectors and spectral norm for symmetric matrices. For a symmetric matrix $\Pm$, $\lambda_i(\Pm)$, $i=1, \ldots, N$, denotes the $i^{\rm th}$ smallest eigenvalue, $\onevect: = [ 1 \; 1 \ldots 1]^{\rm T}$, and $\I$ denotes the identity matrix. 

\section{Review of Network Graph Theory} \label{sec:review_spectral_robust}
In this section, we provide a brief background on network graph theory. Consider an undirected graph $\mathbb{G}=(\mathbb{N}, \mathbb{E})$ containing a set of nodes $\mathbb{N}=\{1, \ldots, N\}$ and a set of edges $\mathbb{E}$. Nodes that communicate with each other have an edge between them. We denote the set of neighbours of node $i$ by $\mathbb{N}_{i}$, $\mathbb{N}_{i}=\{j|\{i,j\} \in \mathbb{E}\}$ where $\{i,j\}$ indicates an edge between the nodes $i$ and $j$ \cite{chung}. A graph is connected if there exists at least one path between every pair of nodes. We denote the number of neighbours of a node $i$ by $d_i$ and $d_{\rm max}=\max_{i} d_i$. The graph structure is described by an $N \times N$ symmetric matrix called the adjacency matrix $\A$, whose $i,j$ element $[\A]_{i,j}=1$ if $\{i,j\} \in \mathbb{E}$. The diagonal matrix $\D ={\rm diag} [d_{1},\;  d_{2}, \; \ldots, \; d_{N}]$ captures the degrees of all the nodes in the network. The Laplacian matrix of the graph is defined as $\La:=\D - \A$. The graph Laplacian characterises a number of useful properties of the graph. The eigenvalues of $\La$ are non-negative and the number of zero eigenvalues denotes the number of distinct components of the graph. When the graph is connected, $\lambda_1 (\La) =0$, and $\lambda_i (\La) > 0 , i \geq 2$,  so that the rank of $\La$ for a connected graph is $N-1$. The vector $\onevect$ is the eigenvector of $\La$ associated with the eigenvalue $0$, i.e, $\La \onevect =\mathbf{0}$. The eigenvalue $\lambda_2 (\La)$ characterizes how densely the graph is connected and the performance of consensus algorithms depend on this eigenvalue \cite{OlfatiSaber2004}.

\section{Sensing and Channel Model} \label{sec:consensus_no_noise_robust}
\subsection{Sensing Model}\label{subsec:nld_sys_model_robust}
Consider a WSN with $N$ sensor nodes each with an initial measurement $x_i(0) \in \R$, $i = 1, \ldots, N$. Measurements made at the sensor nodes are modeled as
\begin{equation}\label{eqn:sensing_model_consensus_robust} 
x_i (0) = \theta + \eta_i  \;, \hspace{0.2 in} i = 1, \ldots, N
\end{equation}
where $\theta$ is an unknown real-valued parameter and $\eta_i$ is the sensing noise at the $i^{\rm th}$ sensor. For many distributions on $\eta_i$, the sample mean of these initial measurements is the maximum likelihood estimate of $\theta$:
\begin{equation}
\label{eq:sample_avg_robust}
\xbar= \frac{1}{N} \displaystyle\sum_{i=1}^{N} x_{i}(0) \;.
\end{equation}
We would like to design an iterative distributed algorithm, in which each sensor communicates only with its neighbours and each sensor has a state that converges to $\xbar$. If the states of all the sensor nodes converge to  $\xbar$, then the network is said to have reached \emph{consensus} on the sample average.  

\subsection{Channel Model}\label{subsec:nld_chn_model_robust}

Each sensor can transmit or receive information to or from its neighbours. When a sensor transmits its state information, it can send a function of its state instead of the state itself. In this link there is additive noise at the receiver node which can be modeled as

\begin{equation}
\label{eq:nld_ch_noise_recvd_info}
y_{ij}(t) = h(x_{j}(t)) + n_{ij}(t) , \; \{i,j\} \in \mathbb{E} \;,
\end{equation}
where $x_{j}(t), j \in \Ni$, is the state value of the $j^{\rm th}$ node at time $t$;  $h(\cdot): \R \rightarrow \R$ is the power-constraining transmission function used at every node, $n_{ij}(t)$ is the noise associated with the reception of $h(x_{j}(t))$, and $y_{ij}(t)$ is the received signal at node $i$ from node $j$ at time $t$. The existing linear consensus algorithms in \cite{Boyd2007,Touri2009,MinyiHuang2007,Pescosolido2008,Barbarossa2008,MinyiHuang2008,AysalBarner2010,Nedic2011cvx, KarMoura2009,KarMoura2007} require $n_{ij}(t)$ to have finite moments. Instead, we assume that the noise samples $n_{ij}(t)$ are mutually independent identically distributed (i.i.d.), symmetric real-valued with zero median (e.g., its PDF, when it exists, is symmetric about zero).

\section{Robust Consensus with Impulsive Communication Noise} \label{sec:consensus_with_noise_robust}
In this section, we propose a robust consensus algorithm in which every node performs a nonlinear operation by soft-limiting the noisy state information at the receiver node. The receiver non-linearity makes the algorithm robust to a wide range of heavy-tailed channel noise distributions. Also, at the transmitter side every sensor maps its state value through a bounded function before transmission to constrain the transmit power making it ideal for resource-constrained WSNs. 

\subsection{The $\rnld$ Algorithm with Communication Noise}\label{subsec:nld_with_noise_robust}
As discussed in \eqref{eq:nld_ch_noise_recvd_info}, each sensor maps its state value at time $t$ through the function $h(x)$ before transmission, and combines the received state values through a nonlinear function $f(x)$ according to the following recursion: 
\begin{align}
\nonumber
x_{i}(t+1) & = x_{i}(t) - \alpha(t) \displaystyle\sum_{j \in \mathbb{N}_{i} } \left [ f \left( h(x_{i}(t)) - y_{ij}(t) \right) \right ] \;,\\                
\label{eq:nld_ch_noise_robust}
           & = x_{i}(t) - \alpha(t) \displaystyle\sum_{j \in \mathbb{N}_{i} } \left [ f \left( h(x_{i}(t)) - h(x_{j}(t)) - n_{ij}(t) \right) \right ] \;,
\end{align}
where $i=1, \ldots, N$, and $t=0,1,2,\ldots$, is the time index, and $\alpha(t)$ is a positive step size which will be assumed to satisfy assumption \textbf{(A5)} in the sequel. The node $j$ transmits its information $x_{j}(t)$ by mapping it through the function $h(x)$, node $i$ receives a noisy signal $h(x_{j}(t))+ n_{ij}(t)$. The function $f(x)$ is applied at the receiver side to combat the effect of impulsive channel noise $n_{ij}(t)$ and will be further assumed to satisfy \textbf{(A2)} in the sequel.

We now compare the existing work on nonlinear consensus in \cite{KhanKar,OlfatiSaber2003,Ulrich2008,HuiWassim2008,WenwuChen2011,Ajorlou2011} against the proposed algorithm in \eqref{eq:nld_ch_noise_robust}. The algorithm in \cite{KhanKar} becomes a special case of \eqref{eq:nld_ch_noise_robust} with $h(x)=x$ and $f(x)=\sin(x)$ in a setting with no channel noise ($n_{ij}(t) \equiv 0$). The algorithm in \cite{OlfatiSaber2003} becomes a special case of \eqref{eq:nld_ch_noise_robust} with $h(x)=x$ and $f(x)$ being an increasing odd function. There is no communication noise assumed in all the existing work on consensus with nonlinear $f(\cdot)$ \cite{KhanKar,OlfatiSaber2003,Ulrich2008,HuiWassim2008,WenwuChen2011,Ajorlou2011} whereas we consider herein the communication noise in the presence of both the transmit and receive non-linearities. Moreover, with the transmit non-linearity $h(x)$, the transmit power from all the sensors are always bounded which is a desirable feature for power constrained WSNs. The $\nld$ algorithm considered in \cite{dastep2013} is a special case of \eqref{eq:nld_ch_noise_robust} with $f(x)=x$ but assumes noise samples have finite moments, and fails in the presence of impulsive channel noise.

We make the following assumptions on $f(x)$, $h(x)$, $n_{ij}(t)$, $\alpha(t)$ and the graph: 
\\
\textbf{Assumptions \;}\\
\textbf{(A1) Graph:\;} The graph $\G$ is undirected and connected so that $\lambda_2 (\La) > 0$ \cite{chung}.\\
\textbf{(A2) Receive Nonlinearity:\;} The function $f(x)$ is strictly increasing, odd and bounded.\\   
\textbf{(A3) Transmit Nonlinearity:\;} The function $h(x)$ is strictly increasing.\\ 
\textbf{(A4) Independent Noise Sequence:\;} The noise samples $n_{ij}(t)$ are mutually i.i.d., symmetric real-valued with zero median (e.g., its PDF, when it exists, is symmetric about zero). \\  
\textbf{(A5) Decreasing Weight Sequence:\;} In order to control the variance growth rate of the cumulative noise we need the following standard conditions on the sequence $\alpha(t)$:
\begin{equation}
\label{eq:assump_A6}
\alpha(t) >0 \;, \; \displaystyle\sum_{t=0}^{\infty} \alpha(t) =  \infty \;,  \; \displaystyle\sum_{t=0}^{\infty} \alpha^{2}(t) < \infty \;.
\end{equation} 

Let $g(x): \R \rightarrow \R$ be such that $g(x) := \E_{n}\left[ f( x + n ) \right]$ where $\E_{n}[\cdot]$ denotes the expectation with respect to any of the i.i.d. $n_{ij}(t)$ so that $f( x + n ) = g(x) + v(x,n)$. Here $v(x,n) = f( x + n ) - \E_{n}\left[ f( x + n ) \right]$ is a noise process which depends on $x \in \R$ and its randomness is due to the noise process $n$, and satisfies $\E_{n} [v(x,n)]=0, x \in \R$. Let $\sigma^2:=\sup_{x} {\rm var}[ f(x+n )]$. Since $f(\cdot)$ is bounded due to \textbf{(A2)}, $\sigma^2$ is finite. Hence we have ${\rm var}[f( x + n )]= {\rm var}[v(x,n)]=\E [v^2(x,n)]\leq \sigma^2$. Using the fact that $f(x)$ is a strictly increasing odd function and that $-n$ has the same distribution as $n$ due to symmetry, it can be easily proved that $g(x)$ is a strictly increasing odd function satisfying $g(0)=0$. Using $g(x)$, the recursion in \eqref{eq:nld_ch_noise_robust} can be written as

\begin{equation}
\label{eq:nld_ch_noise_robust_g_x}
 x_{i}(t+1) = x_{i}(t) - \alpha(t) \displaystyle\sum_{j \in \mathbb{N}_{i} } \left [ g(h(x_i(t))-h(x_{j}(t)))  +  v( h(x_i(t))-h(x_{j}(t)), n_{ij}(t) ) \right ] \;.
\end{equation}
The recursion in \eqref{eq:nld_ch_noise_robust_g_x} can be written in vector form as
\begin{equation} 
\label{eq:nld_vector_ch_noise_robust}
\X(t+1) = \X(t) - \alpha(t) \left [ \mXt + \nv(t, \X(t)) \right ]\;,
\end{equation}
where $\mathbf{X}(t) \in \Rn$ is the state vector at time $t$ given by $\mathbf{X}(t)=[x_{1}(t) \; x_{2}(t) \; \ldots \; x_{N}(t)]^{\rm T}$, and $\boldsymbol\mu(\x): \Rn \rightarrow \Rn$ is a function with $i^{th}$ element is given by
\begin{equation}
\label{eq:nld_mu_comp_robust}
[\boldsymbol\mu(\x)]_{i} =  \displaystyle\sum_{j \in \mathbb{N}_{i}} g(h(x_i)-h(x_{j}))  \;, \itN \;,
\end{equation}
and $\xX=[x_{1} \; x_{2} \; \ldots \; x_{N}]^{\rm T}$. Due to the fact that $g(x)$ is odd and that the graph is connected, we have $\onevectT \mx=0$. The vector $\nv(t,\X(t))$ in \eqref{eq:nld_vector_ch_noise_robust} captures the additive noise at $N$ nodes contributed by their neighbours and their state values and its $i^{th}$ component is given by
\begin{equation} 
\label{eq:nld_ch_noise_comp_robust}
[\nv(t,\X(t))]_{i}=  - \displaystyle\sum_{j \in \mathbb{N}_{i}} v( h(x_i(t))-h(x_{j}(t)), n_{ij}(t) )  \;, \itN \;.
\end{equation}
Clearly, conditioned on $\X(t)=\x$, the noise $\{ v( h(x_i)-h(x_{j}), n_{ij}(t) ) \}_{t\geq 0, 1\leq i,j \leq N}$ is an independent sequence across time $t$, and sensors $i$ due to assumption \textbf{(A4)}. It also satisfies
\begin{align}
\label{eq:assump_A44_robust}
\E [\nv(t, \x)]  = \mathbf{0} \;, \forall t, \x \;, \; \; \vrho:= \sup_{t,\x} \E [\| \nv(t, \x) \|^2] \leq N d_{\rm max} \sigma^2 < \infty.
\end{align}
Note that the inequality in \eqref{eq:assump_A44_robust} is because of \textbf{(A2)} and the fact that the number of neighbours of a given node is upper bounded by $d_{\rm max}$.

We will prove convergence of the $\rnld$ algorithm in Section \ref{subsec:conv_res_dmp_robust} and asymptotic normality in Section \ref{subsec:asym_norm_nld_robust}. We now present a result on the convergence of a discrete time Markov process which will be used in establishing convergence of the $\rnld$ algorithm. 

\subsection{A Result on the Convergence of Discrete time Markov Processes}\label{subsec:conv_res_dmp_robust}
Let $\dmp=\{\X(t)\}_{t \geq 0}$ be a discrete time vector Markov process on $\Rn$. The generating operator $\Lg$ of $\dmp$ is defined as
\begin{equation} %
\label{eq:dmp_generator_robust}
\Lg \vX = \E \left[ \vecXtone |\X(t)=\xX \right] - \vX \;
\end{equation}
for functions $\vX : \Rn \rightarrow \R$, for which the conditional expectation exists. Let $\Bs \subset \Rn$ and its complement be $\Cprime = \Rn \setminus \Bs$. We now state the desired result as a simplification of Theorem 2.7.1 in \cite{Nevelson1973} (see also Theorem 1 in \cite{KarMoura2009}). In general $\Lg \vX$ may depend on $t$. 

\begin{thm} \label{nld_conv_dmp_res_thm_robust}
Let $\dmp$ be a discrete time vector Markov process with the generator operator $\Lg$ as in \eqref{eq:dmp_generator_robust}. If there exists a potential function $\vX : \Rn \rightarrow \R^{+}$, and $\Bs \subset \Rn$ with the following properties
\begin{align}
\label{eq:nld_conv_dmp_res_thm1_robust}
 \vX  > 0,  \xX \in \Cprime, \;\; \vX  = 0,   \;  \xX \in \Bs \;,
\end{align}
\begin{equation} %
\label{eq:nld_conv_dmp_res_thm4_robust}
\Lg \vX \leq - \gamma(t) \fitX + m \zeta(t) [1+ \vX]
\end{equation}
where $m>0$, $\fitX$ is such that
\begin{align}
\label{eq:nld_conv_dmp_res_thm4a_robust}
\fitX =0, \xX \in \Bs, \;  \fitX  > 0, \xX \in \Cprime \;,
\end{align}
and
\begin{align}
\label{eq:nld_conv_dmp_res_thm6_robust}
\gamma(t)  > 0, \zeta(t)  > 0, \; \displaystyle\sum_{t=0}^{\infty} \gamma(t) =  \infty, \; \displaystyle\sum_{t=0}^{\infty} \zeta(t) < \infty \;,
\end{align}
then, the discrete time vector Markov process $\dmp=\{\X(t)\}_{t \geq 0}$ with arbitrary initial distribution converges almost surely (a.s.) to the set $\Bs$ as $t \rightarrow \infty$. That is,

\begin{equation} %
\label{eq:nld_conv_dmp_res_thm8_robust}
{\rm Pr}\left[ \lim_{t \rightarrow \infty} \inf_{\Z \in \Bs} \; \| \X(t) - \Z \| =0 \right]=1.
\end{equation}
\end{thm}

Intuitively, Theorem \ref{nld_conv_dmp_res_thm_robust} indicates that if the one-step prediction error of the Markov process evaluated at the potential function in \eqref{eq:dmp_generator_robust} is bounded as in \eqref{eq:nld_conv_dmp_res_thm4_robust} then it is possible to establish convergence of $\X(t)$.

To prove the a.s. convergence of the consensus algorithm in \eqref{eq:nld_vector_ch_noise_robust} using Theorem \ref{nld_conv_dmp_res_thm_robust}, we choose the consensus subspace $\Cs$, the set of all vectors whose entries are of equal value as,

\begin{equation} %
\label{eq:consensus_subspace}
\Cs=\{ \xX \in \Rn | \xX = a \onevect \;, a \in \R \} \;.
\end{equation}
We are now ready to state the main result of Section \ref{sec:consensus_with_noise_robust}. But first, we start out with a preparatory lemma.

\begin{lem} \label{nld_lem_XMmx_positive} 
Define a positive semi-definite matrix $\Pm$ as the Laplacian of a fully connected graph: $\Pm:=N \I - \onevect \onevectT$. Let $\xX \in \Cprime$, then $\xX^{\rm T} \Pm \mx >0$.
\end{lem}
\begin{IEEEproof}
Consider
\begin{align}
\xX^{\rm T} \Pm \mx     &  =  \xX^{\rm T}[N \I - \onevect \onevectT ] \mx \;,\\
\label{eq:nld_thm_fitX_greater_than_zero3a_robust}
					    &  =  N \xX^{\rm T} \mx - \xX^{\rm T} \onevect \onevectT \mx \;,\\
\label{eq:nld_thm_fitX_greater_than_zero4_robust}
     &  =  N \xX^{\rm T} \mx \;,
\end{align}
where we have used the fact that $\onevectT \mx=0$ in \eqref{eq:nld_thm_fitX_greater_than_zero3a_robust} to get \eqref{eq:nld_thm_fitX_greater_than_zero4_robust}. Expanding $\xX^{\rm T} \mx$ using \eqref{eq:nld_mu_comp_robust}, we get 
\begin{align} 
\nonumber
\label{eq:nld_thm_fitX_greater_than_zero6_robust} 
\xX^{\rm T} \Pm \mx  &  = N \left [ \displaystyle\sum_{j \in \Ns_{1}} g(h(x_1)-h(x_j)) x_1 + \displaystyle\sum_{j \in \Ns_{2}} g(h(x_2)-h(x_j)) x_2 \right. \\
	 & \left. \hspace{0.5 in}+ \ldots + \displaystyle\sum_{j \in \Ns_{N}} g(h(x_N)-h(x_j)) x_N \right ]\;. 
\end{align}
Note that the $i^{\rm th}$ summation in \eqref{eq:nld_thm_fitX_greater_than_zero6_robust} corresponds to the $i^{\rm th}$ node. Now suppose that node $i$ is connected to node $j$. Then there exists a term $g(h(x_i)-h(x_j)) x_i$ in the summation corresponding to the $i^{\rm th}$ node in \eqref{eq:nld_thm_fitX_greater_than_zero6_robust}, and a term $g(h(x_j)-h(x_i)) x_j$ in the summation corresponding to the $j^{\rm th}$ node in \eqref{eq:nld_thm_fitX_greater_than_zero6_robust}. Both of these terms can be combined as $(x_i-x_j) g(h(x_i)-h(x_j))$ and this corresponds to the edge $\{i,j\} \in \mathbb{E}$. Thus equation \eqref{eq:nld_thm_fitX_greater_than_zero6_robust} can be written as pairwise products enumerated over all the edges in the graph as follows
\begin{align} 
\label{eq:nld_thm_fitX_greater_than_zero7_robust}    
\xX^{\rm T} \Pm \mx & = N \displaystyle \sum_{\{i,j\} \in \mathbb{E}} (x_i-x_j) \; g(h(x_i)-h(x_j)) \;.
\end{align}
Since $\xX \in \Cprime$, $\fitX$ in \eqref{eq:nld_thm_fitX_greater_than_zero7_robust} is positive due to the facts that $h(x)$ is strictly increasing and $g(x)$ is a strictly increasing odd function so that there is at least one term in the sum which is greater than zero and this completes the proof. 
\end{IEEEproof}

\begin{thm} \label{nld_thm_as_conv_fixed_graph_robust} 
Let the assumptions \textbf{(A1)}-\textbf{(A5)} hold. Consider the $\rnld$ algorithm in \eqref{eq:nld_vector_ch_noise_robust} with the initial state vector $\X(0) \in \Rn$. Then, the state vector $\X(t)$ in \eqref{eq:nld_vector_ch_noise_robust} approaches the consensus subspace $\Cs$ a.s., i.e.,

\begin{equation} %
\label{eq:nld_thm_as_conv_fixed_graph_robust}
{\rm Pr}\left[ \lim_{t \rightarrow \infty} \inf_{\Z \in \Cs} \; \| \X(t) - \Z \| =0 \right]=1.
\end{equation}
\end{thm}

\begin{IEEEproof}
We will make use of Theorem \ref{nld_conv_dmp_res_thm_robust} to prove \eqref{eq:nld_thm_as_conv_fixed_graph_robust}. We will choose an appropriate potential function $\vX$ that is non-negative and satisfies equation \eqref{eq:nld_conv_dmp_res_thm1_robust}. We will then prove that the generating operator $\Lg$ applied on $\vX$ as in \eqref{eq:dmp_generator_robust} can be upper bounded as in \eqref{eq:nld_conv_dmp_res_thm4_robust} with $\gamma(t)=\alpha(t)$, $\zeta(t)= \alpha^2(t)$, and a $\fitX$ will be chosen to satisfy \eqref{eq:nld_conv_dmp_res_thm4a_robust}. 

First we see that under the assumptions the discrete time vector process $\{\X(t)\}_{t \geq 0}$ in \eqref{eq:nld_vector_ch_noise_robust} is Markov. Let $\Pm$ be a positive semi-definite matrix as defined in Lemma \ref{nld_lem_XMmx_positive}. Let $\vX=\XPX$, then the function $\vX$ is non-negative since $\Pm$ is a positive semi-definite matrix. Note that any $\xX \in \Cs$ is an eigenvector of $\Pm$ associated with the zero eigenvalue, therefore we have
\begin{align}
\label{eq:nld_thm_as_conv_fixed_graph2_robust}
\vX  = 0, \xX \in \Cs \;.
\end{align}  
We have now verified that $\vX$ satisfies the second condition in \eqref{eq:nld_conv_dmp_res_thm1_robust}. We now proceed to show the first condition. Let $\xX = \xX_{\Ss} + \xX_{\Ss \perp}$ where $\xX_{\Ss}$ is the orthogonal projection of $\xX$ on $\Ss$. When $\xX \in \Cprime$, we have $\| \Xcp \| > 0$. Therefore, for any $\xX \in \Cprime$,
\begin{align}
\label{eq:nld_thm_as_conv_fixed_graph3_robust}
\vX   = V(\xX_{\Ss}) + V(\xX_{\Ss \perp}) = V(\xX_{\Ss \perp}) \geq \lambda_2 (\Pm) \| \Xcp \|^2 >0 \;,
\end{align}
where the last inequality is due to $\lambda_2 (\Pm) > 0$. The equations \eqref{eq:nld_thm_as_conv_fixed_graph2_robust} and \eqref{eq:nld_thm_as_conv_fixed_graph3_robust} establish that the conditions in \eqref{eq:nld_conv_dmp_res_thm1_robust} in Theorem \ref{nld_conv_dmp_res_thm_robust} are satisfied.

Let $\xX \in \Cprime$ and $\mx$ be as defined in \eqref{eq:nld_mu_comp_robust}, and $\mXc$ be the orthogonal projection of $\mx$ on $\Ss$. Then, $\mx = \mXc + \mXcp$, where $\mXcp$ is non-zero, i.e., $\| \mXcp\| > 0$ which is proved now. First we recall that $\xX^{\rm T} \Pm \mx >0$ when $\xX \in \Cprime$ due to Lemma \ref{nld_lem_XMmx_positive}. This means $(\xX_{\Ss} + \xX_{\Ss \perp}) \Pm (\mXc + \mXcp)$ = $\xX_{\Ss \perp} \Pm \mXcp >0$ for $\xX \in \Cprime$. If $\mXcp$ were zero, then $\xX_{\Ss \perp} \Pm \mXcp =0$ which contradicts with the fact that $\xX_{\Ss \perp} \Pm \mXcp > 0$. Therefore, $\mXcp$ is non-zero. Define $\beta := \sup_{\xX} \| \mXcp\|^2 / \| \Xcp\|^2 $, then $ 0< \beta < \infty$, where the finiteness of $\beta$ can be seen from the fact that $\mx$ is bounded for all $\xX$ because $f(x)$ is bounded due to $\textbf{(A2)}$, and by expressing $\mx$ around $\xX=a \onevect, a \in \R$ using Taylor's series and observing that the ratio $\| \mx \|^2 / \| \Xcp\|^2$ is finite as $\xX \rightarrow a \onevect$. 

Now we will prove that \eqref{eq:nld_conv_dmp_res_thm4_robust} is satisfied as well. Towards this end, consider $\Lg \vX $ defined in \eqref{eq:dmp_generator_robust},
\begin{align}
\label{eq:nld_conv_dmp_res_thm7a_robust}
\Lg \vX &= \E \left[ \XTMXt |\X(t)=\xX \right] - \vX \;, \\ 
\nonumber
		&= \E \left[ \left( \xX^{\rm T} - \alpha(t) \left( \mx^{\rm T} + \nv^{\rm T}(t,\xX) \right) \right) \cdot   \left(\Pm \xX - \alpha(t) \left( \Pm \mx + \Pm \nv(t,\xX)  \right) \right)  \right] \\
\label{eq:nld_conv_dmp_res_thm8a_robust}
		& \;\;\;\;\;\; - \vX \;, \\
\label{eq:nld_conv_dmp_res_thm9_robust}
		&= - 2 \alpha(t) \xX^{\rm T} \Pm \mx  + \alpha^2(t) \mx^{\rm T} \Pm \mx  + \E  \left[ \nv^{\rm T}(t,\xX) \Pm \nv(t,\xX)  \right] .
\end{align}
We get \eqref{eq:nld_conv_dmp_res_thm9_robust} by expanding \eqref{eq:nld_conv_dmp_res_thm8a_robust} and taking the expectations and using the fact that $\E [\nv(t, \xX)]=\mathbf{0}$. We have 
\begin{equation}
\label{eq:nld_thm_as_conv_fixed_graph11_robust}
\E  \left[ \nv^{\rm T}(t,\xX) \Pm \nv(t,\xX)  \right] \leq \E  \left[ \lambda_N (\Pm) \| \nv(t,\xX) \|^2 \right]   \leq \lambda_N (\Pm) \vrho \;,
\end{equation}
where the second inequality follows from \eqref{eq:assump_A44_robust}. Using \eqref{eq:nld_thm_as_conv_fixed_graph11_robust} in \eqref{eq:nld_conv_dmp_res_thm9_robust}, we get the following bound
\begin{align}
\label{eq:nld_thm_as_conv_fixed_graph12_robust}
\Lg \vX & \leq - 2 \alpha(t) \left[ \xX^{\rm T} \Pm \mx \right] + \alpha^2(t) \left[ \mx^{\rm T} \Pm \mx  +  \vrho \lambda_N (\Pm) \right] \;, \\
\label{eq:nld_thm_as_conv_fixed_graph13_robust}
		& \leq - 2 \alpha(t) \left[ \xX^{\rm T} \Pm \mx \right] + \alpha^2(t) \left[  \lambda_N(\Pm) \beta \| \Xcp \|^2  + \vrho \lambda_N (\Pm)  \right] \;, \\
\label{eq:nld_thm_as_conv_fixed_graph14_robust}
		& \leq - 2 \alpha(t) \left[ \xX^{\rm T} \Pm \mx \right] + \alpha^2(t) \left[ \beta  \XPX + \vrho N  \right] \;, \\ 
\label{eq:nld_thm_as_conv_fixed_graph15_robust}
		& \leq - 2 \alpha(t) \left[ \xX^{\rm T} \Pm \mx \right] + m  \alpha^2(t) \left[  1+ \beta_2 \XPX \right] \;, \\	
\label{eq:nld_thm_as_conv_fixed_graph16_robust}
		& \leq - \alpha(t) \fitX + m  \alpha^2(t) \left[ 1+ \vX \right] \;,			
\end{align}
where we have used the fact $\mx^{\rm T} \Pm \mx \leq \lambda_N(\Pm) \| \mXcp \|^2$ and $\| \mXcp \|^2 \leq \beta \| \Xcp \|^2 $ in \eqref{eq:nld_thm_as_conv_fixed_graph12_robust} to get \eqref{eq:nld_thm_as_conv_fixed_graph13_robust}. In \eqref{eq:nld_thm_as_conv_fixed_graph13_robust}, we have used the fact that $\XPX \geq \lambda_2(\Pm) \| \Xcp \|^2$ due to  \eqref{eq:nld_thm_as_conv_fixed_graph3_robust} and $\lambda_2(\Pm)=\lambda_N(\Pm)=N$ to get \eqref{eq:nld_thm_as_conv_fixed_graph14_robust}. In \eqref{eq:nld_thm_as_conv_fixed_graph14_robust}, we defined  $m := \max \{  \beta, \vrho N  \}$, and $\beta_2 := \vrho N / m$ to get \eqref{eq:nld_thm_as_conv_fixed_graph15_robust} and it is easy to see that $\beta_2 \in (0, 1]$. From  \eqref{eq:nld_thm_as_conv_fixed_graph15_robust}, due to the fact that $\beta_2 \in (0, 1]$ and letting $\fitX := 2 \xX^{\rm T} \Pm \mx$, we get \eqref{eq:nld_thm_as_conv_fixed_graph16_robust}.

We will now prove that $\fitX$ in \eqref{eq:nld_thm_as_conv_fixed_graph16_robust} satisfies equation \eqref{eq:nld_conv_dmp_res_thm4a_robust} of Theorem \ref{nld_conv_dmp_res_thm_robust}.

Whenever $\xX \in \Cs$, i.e., $\xX = a \onevect, a \in \R$, then $x_i=x_j, \forall i,j$, which means $g(h(x_i)-h(x_j))=g(0)=0, \forall i,j$, and hence $\mx = \mathbf{0}$. This implies that $\fitX=0, \forall \xX \in \Cs$. From Lemma \ref{nld_lem_XMmx_positive}, it is immediate that $\fitX = 2 \xX^{\rm T} \Pm \mx > 0$ whenever $\xX \in \Cprime$. 

Letting $\gamma(t)=\alpha(t), \zeta(t)=\alpha^2(t)$ and by assumption \textbf{(A5)}, we see that the sequence $\alpha(t)$ in \eqref{eq:nld_thm_as_conv_fixed_graph16_robust} satisfies \eqref{eq:nld_conv_dmp_res_thm6_robust}. Thus all the conditions of Theorem \ref{nld_conv_dmp_res_thm_robust} are satisfied to yield \eqref{eq:nld_thm_as_conv_fixed_graph_robust}.
\end{IEEEproof}

Theorem \ref{nld_thm_as_conv_fixed_graph_robust} states that the sample paths of $\X(t)$ approach the consensus subspace almost surely. Now, like in \cite{KarMoura2009}, we will prove the convergence of $\X(t)$ to a finite point in $\Cs$ in Theorem \ref{nld_conv_as_limiting_rv_robust}. 

\begin{thm} \label{nld_conv_as_limiting_rv_robust}
Let the assumptions of Theorem \ref{nld_thm_as_conv_fixed_graph_robust} hold. Consider the $\rnld$ algorithm in \eqref{eq:nld_vector_ch_noise_robust} with the initial state $\X(0) \in \Rn$. Then, there exists a finite real random variable $\cval$ such that

\begin{equation} %
\label{eq:nld_conv_as_limiting_rv_robust}
{\rm Pr}\left[ \lim_{t \rightarrow \infty} \X(t) = \cval \onevect \right]=1.
\end{equation}

\end{thm}

\begin{IEEEproof}
Let the average of $\X(t)$ be $\xbar(t)=\onevectT \X(t) / N$. It suffices to show that $\{\xbar(t)\}_{t \geq 0}$ is an $\Ltwo$ bounded martingale. A sequence of random variables $\{y(t)\}_{t \geq 0}$ is called as a martingale if for all $t \geq 0$, $\E\left[ |y(t)| \right] < \infty$ and $\E\left[ y(t+1) \; | \; y(1) \; y(2) \ldots y(t) \right]=y(t)$. The sequence $\{y(t)\}_{t \geq 0}$ is an $\Ltwo$ bounded martingale if $\sup_{t} \E\left[ y^2(t) \right] < \infty$ (see \cite[pp. 110]{David1991}). Since $\onevect \xbar(t) \in \Cs$, Theorem \ref{nld_thm_as_conv_fixed_graph_robust} implies,

\begin{equation} %
\label{eq:nld_conv_as_limiting_rv2_robust}
{\rm Pr}\left[ \lim_{t \rightarrow \infty} \| \X(t) -  \xbar(t) \onevect \| =0 \right]=1 \;,
\end{equation}
where \eqref{eq:nld_conv_as_limiting_rv2_robust} follows from \eqref{eq:nld_thm_as_conv_fixed_graph_robust} since the infimum in \eqref{eq:nld_thm_as_conv_fixed_graph_robust} is achieved by $\Z=\xbar(t) \onevect$. Pre-multiplying \eqref{eq:nld_vector_ch_noise_robust} by $\onevectT / N$ on both sides and noting that $\onevectT \mx = 0, \forall \xX$ due to the symmetric structure of the graph we get, 

\begin{align}
\label{eq:nld_conv_as_limiting_rv3_robust}
\xbar(t+1)  & =  \xbar(t) - \tilde{v}(t) \\
\label{eq:nld_conv_as_limiting_rv4_robust}
    & =  \xbar(0) - \displaystyle\sum_{0 \leq k \leq t} \tilde{v}(k) 
\end{align}
where $\tilde{v}(t) =\alpha(t) \onevectT \nv(t,\X(t)) / N$. From \eqref{eq:assump_A44_robust} it follows that
\begin{align}
\nonumber
\E [\tilde{v}(t)] & =0, \\
\nonumber
\label{eq:nld_conv_as_limiting_rv6_robust}
\displaystyle\sum_{t \geq 0} \E [\tilde{v}^2(t)] &=  \displaystyle\sum_{t \geq 0} \frac{\alpha^2(t)}{N^2} \E \left [ \| \nv(t,\X(t)) \|^2 \right ] \leq  \frac{\vrho}{N^2} \displaystyle\sum_{t \geq 0} \alpha^2(t) < \infty
\end{align}
which implies
\begin{equation} %
\label{eq:nld_conv_as_limiting_rv7_robust}
\E [\xbar^2(t+1)] \leq \xbar^2(0) + \frac{\vrho}{N^2} \displaystyle\sum_{t \geq 0} \alpha^2(t)\;, \forall t \;.
\end{equation}
Equation \eqref{eq:nld_conv_as_limiting_rv7_robust} together with \eqref{eq:nld_conv_as_limiting_rv3_robust} implies that the sequence $\{\xbar(t)\}_{t \geq 0}$ is an $\Ltwo$ bounded martingale and hence converges a.s. to a finite random variable $\cval$ (see \cite[Theorem 2.6.1]{Nevelson1973}). Therefore the theorem follows from \eqref{eq:nld_conv_as_limiting_rv2_robust}.
\end{IEEEproof}

In what follows, we present the properties of the limiting random variable $\cval$.

\subsection{Mean Square Error of $\rnld$ Algorithm}\label{subsec:mse_nld_robust}
Theorems \ref{nld_thm_as_conv_fixed_graph_robust} and \ref{nld_conv_as_limiting_rv_robust} establish that the sensors reach consensus asymptotically and converge a.s. to a finite random variable  $\cval$. We can view $\cval$ as an estimate of  $\xbar$. In the following theorem we characterize the bias and mean squared error (MSE) properties of $\cval$.  We define the MSE of $\cval$ as $\mse = \E [(\cval - \xbar)^2].$

\begin{thm} \label{nld_limiting_rv_mse_robust}
Let $\cval$ be the limiting random variable as in Theorem  \ref{nld_conv_as_limiting_rv_robust}. Then $\cval$ is unbiased, $\E [\cval] =\xbar$, and its MSE is bounded, $\mse \leq  \vrho N^{-2} \displaystyle\sum_{t \geq 0} \alpha^2(t)$.
\end{thm}
The proof is obtained by following the same steps of the Lemma 5 in \cite{KarMoura2009}. 

We point out that with non-linear processing at both the transmitter and receiver nodes, we have obtained a similar bound on the MSE $\mse$ as that of the linear consensus algorithm in \cite{KarMoura2009} but in our case the bound depends on the function $f(x)$ (see assumption \textbf{(A2)}) through $\vrho$ but does not depend on $h(x)$. Recall that $\vrho \leq N d_{\rm max} \sigma^2$ from \eqref{eq:assump_A44_robust} which implies that $\mse \leq  d_{\rm max} N^{-1} \sum_{t \geq 0} \alpha^2(t) \sigma^2$. Therefore, if $d_{\rm max}$ is finite for a large connected network, we have $\lim_{N \rightarrow \infty} \mse =0$ and this means that $\cval$ converges to $\xbar$ for large $N$. If the graph is densely connected, then $d_{\rm max}$ is relatively high which increases the worst-case MSE. On the other hand, when the graph is densely connected, $\lambda_2 (\La)$ is larger which aids in the speed of convergence to $\cval$, as quantified through the covariance matrix in Section \ref{subsec:asym_norm_nld_robust}. 

\subsection{Asymptotic Normality of $\rnld$ Algorithm}\label{subsec:asym_norm_nld_robust} 

In this section, we establish the asymptotic normality of the $\rnld$ algorithm in \eqref{eq:nld_vector_ch_noise_robust}. Our approach here is similar to the one in \cite{MinyiHuang2008} and \cite{dastep2013}. Basically, we decompose the $\rnld$ algorithm in $\Rn$ into a scalar recursion and a recursion in $\Rnm$. We now formally state and prove the result as a theorem. 

\begin{thm} \label{asym_norm_nld_lemma_robust}
Let $\alpha(t)=a / (t+1), a>0$, then the $\rnld$ algorithm in \eqref{eq:nld_vector_ch_noise_robust} becomes
\begin{equation} 
\label{eq:asym_norm_nld_lemma1_robust}
\X(t+1) = \X(t) + \frac{a}{t+1} \left [ - \mXt + \nv(t,\X(t)) \right ]. 
\end{equation}
Suppose that the assumptions \textbf{(A1)}-\textbf{(A5)} hold, and that the functions $f(x)$ and $h(x)$ are differentiable with $0 < h^{'}(x)  \leq c$, for some $c>0$. Let the eigenvalue decomposition of $\La$ be given by $\La=\U \Sgma \UT$, where $\U$ is a unitary matrix whose columns are the eigenvectors of $\La$ such that 
\begin{equation} %
\label{eq:asymp_norm_linear2_robust} 
\U=\left [ \frac{\onevect}{\sqrt{N}} \;\; \fiN \right ], \fiN \in \Rmn \;, \; -\Sgma = \begin{bmatrix} 0  &  \mathbf{0}^{\rm T}  \\ \mathbf{0}  &  \B \\ \end{bmatrix} \;,
\end{equation} 
where $\B \in \Rnmm$ is a stable diagonal matrix containing the $N-1$ negative eigenvalues of $-\La$ along its diagonal. In addition, let $\theta_0$ be a realization of the random variable $\cval$ and $a$ is chosen such that $ 2 a \lambda_2(\La) g^{'}(0) h^{'}(\theta_0) >1$ so that the matrix $\left [a g^{'}(0) h^{'}(\theta_0) \B + \I /2 \right ] , \theta_0 \in \R$ is stable. Define $[\tilde{n}(t) \; \; \ntildet^{\rm T}]^{\rm T}:= N^{-1/2}  \UT \nv(t,\X(t)), \; \ntildet \in \Rnm,$ so that $\tilde{n}(t)= N^{-1}  \onevectT \nv(t,\X(t))$ and $\ntildet= N^{-1/2}  \fiNT \nv(t,\X(t))$. Let $\sigma^2_{n}:=\lim_{t \rightarrow \infty} {\rm var}[\tilde{n}(t)]$ and $\Czero := \lim_{t \rightarrow \infty} \E[\ntilde(t) \ntilde(t)^{\rm T} ]$, $\Czero \in \Rnmm$. Then, as $t \rightarrow \infty$,

\begin{equation} %
\label{eq:asym_norm_nld_lemma4_robust}
\sqrt{t}(\X(t) - \theta_0 \onevect) \sim \mathcal{N} \left(0, a^2 \sigma^2_{n} \onevect \onevectT + N^{-1} \fiN \Stheta \fiNT \right) \;,
\end{equation}
where 
\begin{align}
\label{eq:asym_norm_nld_lemma7_robust}
\Stheta    & = a^2 \int\limits_{0}^{\infty} e^{\left [a g^{'}(0) h^{'}(\theta_0) \B + \I/2 \right ] t} \; \Czero \; e^{\left [a g^{'}(0) h^{'}(\theta_0) \B + \I/2 \right ] t} dt \;.
\end{align}
\end{thm}
\begin{IEEEproof}
Define $[\xtilde(t) \;\; \Xtilde(t)^{\rm T} ]^{\rm T}:= N^{-1/2}   \UT \X(t), \Xtilde(t) \in \Rnm$. From Theorem \ref{nld_conv_as_limiting_rv_robust}, we have  $\X(t) \rightarrow \cval \onevect$ a.s. as $t \rightarrow \infty$ which implies that $[\xtildet \; \; \Xtildet ]^{\rm T} \rightarrow [\cval \; \; \mathbf{0} ]^{\rm T}$ a.s. as $t \rightarrow \infty$, and therefore $\Xtildet \rightarrow \mathbf{0}$ a.s. as $t \rightarrow \infty$. For a given $\theta_0$, the error $[\X(t) - \theta_0 \onevect]$ can be written as the sum of two error components (see also Section VI in \cite{MinyiHuang2008}) as given below
\begin{align}
\label{eq:asym_norm_nld_error1_robust}
[\X(t) - \theta_0 \onevect]  & = [\xtildet - \theta_0 ] \onevect + \frac{1}{\sqrt{N}} \fiN \Xtildet \;.
\end{align}
Define $\e_1=[\xtildet - \theta_0 ] \onevect$ and $\e_2= N^{-1/2} \fiN \Xtildet$ as the first and second terms in \eqref{eq:asym_norm_nld_error1_robust}. By calculating the covariance matrix between $\e_1$ and $\e_2$, it can be proved that they are asymptotically uncorrelated as $t \rightarrow \infty$, and that asymptotically $\sqrt{t} \e_1 \sim \mathcal{N} (0, a^2 \sigma^2_{n} \onevect \onevectT)$ (see Theorem 12 in \cite{MinyiHuang2008}) where $\sigma^2_{n}$ is the variance of $\tilde{n}(t)$ as $t \rightarrow \infty$ which is calculated to be $\sigma^2_{n} = ( N^{-2}  \sum_{i=1}^{N} d_i ) \E_n [f^2(n)]$. To show that $\sqrt{t} \e_2$ is asymptotically normal, it suffices to show that  $\sqrt{t} \Xtildet$ is asymptotically normal. To this end, we linearize $\mx$ in \eqref{eq:asym_norm_nld_lemma1_robust} around $\xX=\theta_0 \onevect $ using Taylor's series expansion,

\begin{align} %
\label{eq:asym_norm_nld_proof1_robust}
\mx & = \mathbf{\boldsymbol\mu}(\theta_0 \onevect) + \frac{\partial \mx } {\partial \xX}\bigg|_{ \xX = \theta_0 \onevect} (\xX-\theta_0 \onevect) + o(\|\xX-\theta_0 \onevect\|)\;, \\ 
\label{eq:asym_norm_nld_proof1a_robust}
       & = g^{'}(0) h^{'}(\theta_0) \La \xX + o(\|\xX-\theta_0 \onevect\|)\;, 
\end{align}
where the Jacobian matrix of $\mx$ has $i,j$ element given by 
${\left [\frac{\partial \mathbf{\boldsymbol\mu}(\x)}{\partial \xX} \right]}_{i,j} = \frac{\partial \mathbf{\boldsymbol\mu}_{i}(\x)}{\partial x_{j}}$.

Using \eqref{eq:asym_norm_nld_proof1a_robust} in \eqref{eq:asym_norm_nld_lemma1_robust} we get
\begin{align} %
\label{eq:asym_norm_nld_proof3_robust}
\X(t+1)    & = \X(t) + \frac{a}{t+1} \left [ g^{'}(0) h^{'}(\theta_0) \left( -\La \X(t) \right)  + o(\|\X(t)-\theta_0 \onevect\|)  + \nv(t,\X(t)) \right ]\;, \; {\rm as} \; t \rightarrow \infty.
\end{align}
Pre-multiplying \eqref{eq:asym_norm_nld_proof3_robust} on both sides by $ N^{-1/2} \UT$ and using \eqref{eq:asymp_norm_linear2_robust} we get the following recursions 
\begin{align} 
\label{eq:asym_norm_nld_proof4_robust}
\xtilde(t+1) & = \xtildet + \frac{a}{t+1} \tilde{n}(t)\;, \\
\label{eq:asym_norm_nld_proof5_robust}
\Xtilde(t+1) & = \Xtildet + \frac{a}{t+1} \left [ g^{'}(0) h^{'}(\theta_0) \B \Xtildet + o(\|\X(t)-\theta_0 \onevect\|) + \ntildet \right ], {\rm as} \; t \rightarrow \infty.
\end{align}
In \cite{Nevelson1973}, asymptotic normality  of a recursion similar to \eqref{eq:asym_norm_nld_proof5_robust} has been proved  under certain conditions. With the assumption that $\left [a g^{'}(0) h^{'}(\theta_0) \B + \I / 2 \right ]$ is a stable matrix for $\theta_0 \in \R$, it can be verified that all the conditions of Theorem 6.6.1 in \cite[p. 147]{Nevelson1973} are satisfied for the process $\Xtildet$ in \eqref{eq:asym_norm_nld_proof5_robust}. Therefore, for a given $\theta_0$, the process $\sqrt{t} \Xtildet$ is asymptotically normal with zero mean and covariance matrix given by \eqref{eq:asym_norm_nld_lemma7_robust}. Since $\sqrt{t} \e_1 \sim \mathcal{N} (0, a^2 \sigma^2_{n} \onevect \onevectT)$ and using \eqref{eq:asym_norm_nld_lemma7_robust} together with the fact that $\e_1$ and $\e_2$ are asymptotically independent as $t \rightarrow \infty$, we get \eqref{eq:asym_norm_nld_lemma4_robust} which completes the proof.
\end{IEEEproof}

Equation \eqref{eq:asym_norm_nld_lemma4_robust} indicates how fast the process $\X(t)$ will converge to $\theta_0 \onevect$ for a given $\theta_0$. The convergence speed clearly depends on $g^{'}(0)$ and $h^{'}(\theta_0)$ which captures the effect of receiver and transmit non-linearities respectively. 

Let the asymptotic covariance in \eqref{eq:asym_norm_nld_lemma4_robust} be denoted by $\CSR$. Since $\ntilde(t)$ are asymptotically i.i.d. across space and time, $\Czero$ in \eqref{eq:asym_norm_nld_lemma7_robust} becomes $\Czero= N \sigma^2_{n} \I$ with $\sigma^2_{n} = ( N^{-2}  \sum_{i=1}^{N} d_i ) \E_n [f^2(n)]$ and thus we have $\CSR= a^2 \sigma^2_{n} \onevect \onevectT + N^{-1} \fiN \Stheta \fiNT $ where $\Stheta$ is a diagonal matrix whose diagonal elements are given by $\Stheta_{i,i} = a^2 \sigma^2_{n} / [2 a g^{'}(0) h^{'}(\theta_0) \lambda_{i+1}(\La) -1]$. A reasonable quantitative measure of largeness \cite{Polyak1981} of the asymptotic covariance matrix is  $ \| \CSR \| $ which is the maximum eigenvalue of the symmetric matrix $\CSR$. 

Further, $ \| \CSR \| $ can be minimized with respect to the parameter $a$ and this can be formulated as the following optimization problem,
\begin{equation} 
\label{eq:optimize_asymp_covariance_robust}
\min_{\{a | 2 a g^{'}(0) h^{'}(\theta_0) \lambda_{2}(\La) > 1 \} } \; \max_{\{\x | \x \in \Rn,   \| \x \|^2 \leq 1 \} } \x^{\rm T} \CSR \x \;,
\end{equation} 
which can be solved analytically. The value of $a$ that optimizes \eqref{eq:optimize_asymp_covariance_robust} is found to be $a^{*}_{\rm RC} = (N+1) / [ 2 N \lambda_2(\La) g^{'}(0) h^{'}(\theta_0)]$ and the corresponding optimal value of the $ \| \CSR \| $is given by
\begin{equation} 
\label{eq:asymp_covariance_value_robust}
 \| \CSR^{*} \| = \left ( N^{-2}  \sum_{i=1}^{N} d_i \right) \left (\frac{N+1}{2 N} \right)^2 \left( \frac{\E_n [f^2(n)]}{(\E_n[f^{'}(n)])^2} \right) \left (\frac{1}{\lambda^2_2(\La)} \right )  \left ( \frac{1}{h^{'}(\theta_0)} \right )^2 \;.
\end{equation}
The best speed of convergence characterized by the asymptotic covariance depends on the point of convergence through $h^{'}(\theta_0)$. To select the optimal $a$ that would result in the best speed of convergence for a given $f(x)$ and $h(x)$, knowledge of $\theta_0$ is required. Since $\theta_0$ unknown apriori, the performance characterized in \eqref{eq:asymp_covariance_value_robust} could serve as the benchmark for a given $h(x)$. In practice it may be possible for sensors to adapt the value of $a$ as they converge towards the limiting value $\theta_0$ to speed up the convergence, and approach this benchmark.  An optimized value for $a$, also provides a simpler final expression for the asymptotic covariance in terms of its dependence on the receive nonlinearity $f(x)$ and transmit nonlinearity $h(x)$.

The size of the asymptotic covariance matrix in \eqref{eq:asymp_covariance_value_robust} is inversely proportional to the square of the smallest non-zero eigenvalue $\lambda_2(\La)$ which quantifies how densely a graph is connected. Even though the asymptotic covariance $ \CSR $ has been derived in the literature \cite{MinyiHuang2008}, its optimization has not been considered. The optimization considered in \eqref{eq:optimize_asymp_covariance_robust} enables us to infer some interesting conclusions. In Table \ref{table1}, we have summarized the behavior of $\| \CSR^{*} \|$ for several graphs for large $N$ \cite{deAbreu2007,ZhangDong2011,Olfati-Saber2007,Spielman2012}. For the fully connected graph, $\| \CSR^{*} \|$ goes to zero faster than the star graph and thus the former will converge faster than the latter. For the ring and line graphs, with large $N$, the convergence will become slower since $\| \CSR^{*} \|$ increases with $N$. For other graphs in Table \ref{table1}, the convergence speed is better compared to the line and ring graphs since $\| \CSR^{*} \|$ decreases with $N$ for those graphs. 
\begin{table}[h!]
\begin{center}
\begin{tabular}{|l|l|l|} 
\hline 
\multirow{1}{*} Type of Graph & $\lambda_2(\La)$ & Behavior of $\| \CSR^{*} \|$ \\ \hline
\multirow{1}{*} Fully Connected  & $N$ & $O \left(N^{-2} \right)$  \\ \hline
\multirow{1}{*} Star &  1 & $O \left (N^{-1} \right)$ \\ \hline
\multirow{1}{*} Ring  & 4 $\sin^2 \left(\frac{\pi}{N} \right)$ & $O \left(N^{3}\right)$ \\ \hline
\multirow{1}{*} Line  & 4 $\sin^2 \left(\frac{\pi}{2N} \right)$ & $O \left(N^{3}\right)$ \\ \hline
\multirow{1}{*} Tree (excluding star graphs) & $\leq 0.3819$ & $O \left (N^{-1} \right)$ \\ \hline
\multirow{1}{*} Cubic Graph  & 2 & $O \left (N^{-1} \right)$ \\ \hline
\multirow{1}{*} Planar  & $\leq 4$ & $O \left (N^{-1} \right)$ \\ \hline
\multirow{1}{*} Bipartite complete graph with $p$ and $q$ vertices  & $\min(p,q)$ & $O \left (N^{-1} \right)$ \\ \hline
\multirow{1}{*} k-regular (includes Ramanujan graphs)  & $\leq k - 2 \sqrt{k-1}$ & $O \left (N^{-1} \right)$ \\ \hline
\multirow{1}{*} k-regular Lattice  & $(k+1) - \frac{\sin((k+1) \frac{\pi}{N})}{\sin(\frac{\pi}{N})}$ & $O \left (N^{-1} \left ( (k+1) - \frac{\sin((k+1) \frac{\pi}{N})}{\sin(\frac{\pi}{N})}\right)^{-2} \right)$ \\ \hline
\end{tabular}
\caption{Behavior of $\| \CSR^{*} \|$ for some common graphs}
\label{table1}
\vspace{-0.4 in}
\end{center}
\end{table}
It is also interesting to note that the minimization of \eqref{eq:asymp_covariance_value_robust} with respect to the transmit and receive nonlinearities can be done separately and thus asymptotic covariance is an easier and helpful metric in optimizing the performance. The nonlinear receiver function $f(x)$ for which the ratio $ \E_n [f^2(n)] / (\E_n[f^{'}(n)])^2 $ is smaller will be better in terms of speed of convergence. For example, if $n$ is Laplacian distributed with variance of 2 and if $f(x)=x$, then $\E_n [f^2(n)] / (\E_n[f^{'}(n)])^2 =2$ whereas if we choose $f(x)=\tanh(x)$, we have $\E_n [f^2(n)] / (\E_n[f^{'}(n)])^2 =1.317$ indicating $\tanh(x)$ will perform better than the linear case. This is due to the fact that Laplacian is a heavy tailed distribution and therefore a bounded function such as $\tanh(x)$ curtails the effect of outliers which does not happen when $f(x)$ is linear. Equation \eqref{eq:asymp_covariance_value_robust} also indicates when $h(x)$ is fixed, scaling $f(x)$ does not change the speed of convergence. We will illustrate these findings using simulations in Section \ref{sec:simulations_nld_robust}. 

When $f(x)$ is a bounded function, from equation (8) in \cite{chinesephysics} we have
\begin{equation}
\label{eq:chinesephysics_robust}
\frac{\E_n [f^2(n)]}{(\E_n[f^{'}(n)])^2} \geq \frac{1}{J} \;,
\end{equation}
where $J$ is the Fisher information of $n$ with respect to a location parameter \cite[(8)]{zamir} and thus we see an interesting relationship between the maximum eigenvalue of the asymptotic covariance and the Fisher information. For any $h(x)$, the best choice of $f(x)$ is the one that achieves equality in \eqref{eq:chinesephysics_robust}. For instance, when $n$ is Gaussian, $f(x)=x$ achieves equality in \eqref{eq:chinesephysics_robust} in which case we have  ${\rm var}[f(n)]$ equals the inverse of Fisher information. In addition, when $n$ has finite moments, our $\rnld$ algorithm in \eqref{eq:nld_ch_noise_robust} subsumes the non-linear consensus algorithm discussed in \cite{dastep2013} with $f(x)=x$, and we get the same result as in \eqref{eq:asymp_covariance_value_robust} except $\E_n [f^2(n)]$ is replaced by the noise variance $\sigma^2_{v}$ defined in \cite{dastep2013}. Further, our model subsumes the linear case studied in \cite{MinyiHuang2008} with $f(x)=x$ and $h(x)=x$. 

\section{Simulations} \label{sec:simulations_nld_robust}
In this section, we corroborate our analytical findings through various simulations. In all the simulations presented, the initial samples $x_i(0) \in \R, i=1,2, \ldots, N,$ were generated randomly using Gaussian distribution with a standard deviation equal to 10. The desired global average value is indicated in each of the simulations. We focus here on bounded functions for both the transmit and receiver non-linearities to study their performance. 
 
\subsection{Performance of $\rnld$ algorithm with Channel Noise}\label{subsec:perf_nld_withs_noise_robust}
\begin{figure}[tb]
\begin{minipage}{1\textwidth}
\centering
\begin{center}
\includegraphics[height=9.2cm,width=12cm]{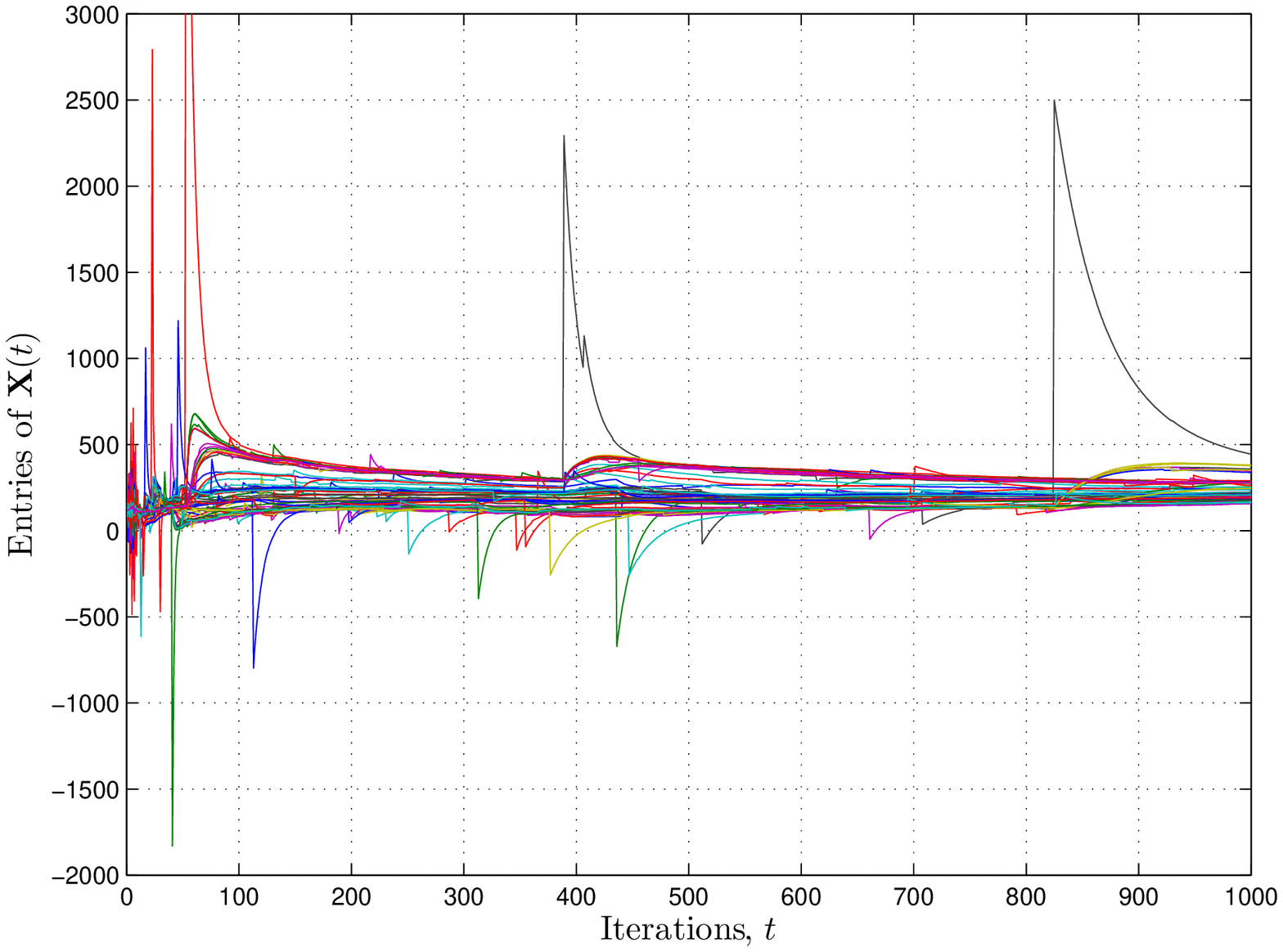} 
\caption{Linear consensus fails with impulsive channel noise, Entries of $\mathbf{X}(t)$ versus Iterations $t$: Cauchy noise, $h(x)= x, f(x)= x$, $N=75$, $\xbar=134.31$, $\gamma=1$.}\label{fig:Linear_Fails_Cauchy}
\end{center}
\end{minipage}
\end{figure}

First, we highlight that the linear consensus algorithms in \cite{Boyd2007,Touri2009,MinyiHuang2007,Pescosolido2008,Barbarossa2008,MinyiHuang2008,AysalBarner2010,Nedic2011cvx, KarMoura2009,KarMoura2007} fail to achieve consensus when the channel noise does not have finite variance. An example plot is shown in Figure \ref{fig:Linear_Fails_Cauchy} for the case when the channel noise is Cauchy distributed with the scale parameter $\gamma=1$. Clearly, the sensors do not reach consensus. Whereas the proposed $\rnld$ algorithm works when we choose $f(x)$ as a nonlinear function as shown next. 

\begin{figure}[tb]
\begin{minipage}{1\textwidth}
\centering
\begin{center}
\includegraphics[height=9.2cm,width=12cm]{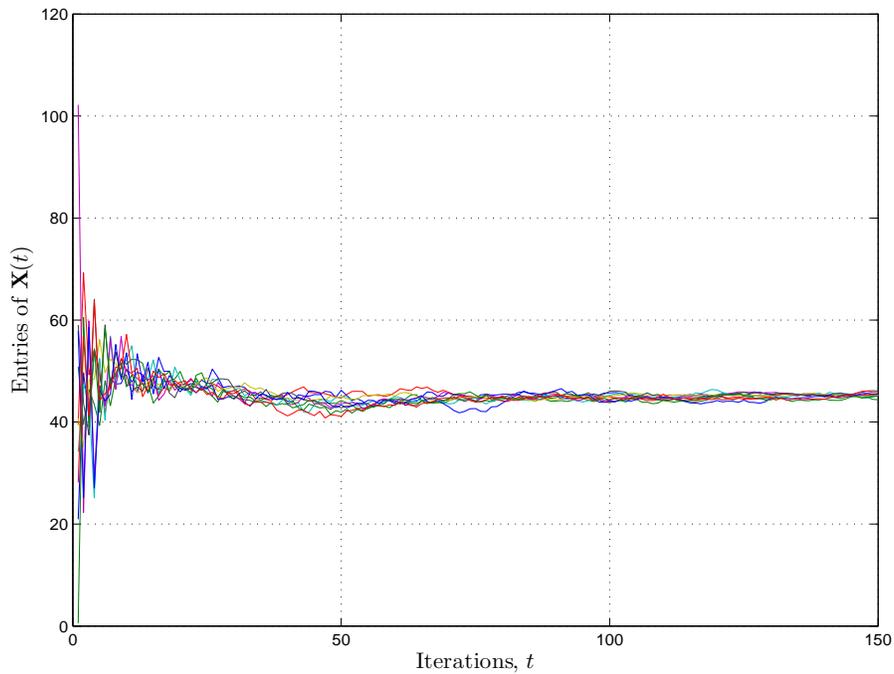} 
\caption{Convergence of $\rnld$ algorithm for a small Graph, Entries of $\mathbf{X}(t)$ versus Iterations $t$: Cauchy noise, $h(x)= \sqrt{\rho} \frac{2}{\pi} \tan^{-1}(\frac{\pi}{2} 0.01 x), f(x)= \tanh(5 x)$, $N=10$, $\xbar=43.96$, $\rho=15$ dB, $\gamma=0.1$.}\label{fig:TxRx_Cauchy_TanxArcTanx_SmallWSN_Noisy}
\end{center}
\end{minipage}
\end{figure}

\begin{figure}[tb]
\begin{minipage}{1\textwidth}
\centering
\begin{center}
\includegraphics[height=9.15cm,width=12cm]{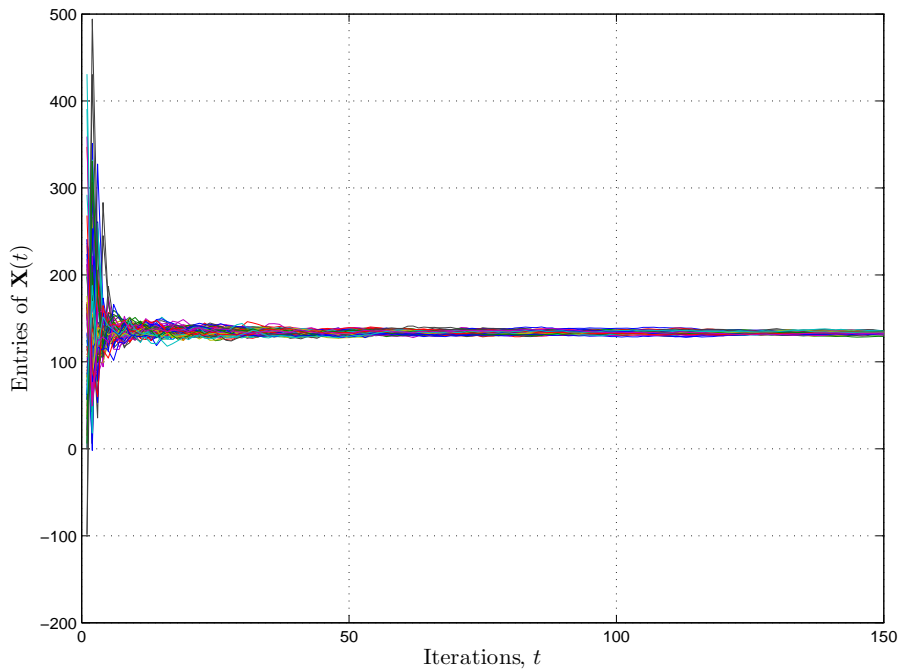} 
\caption{Convergence of $\rnld$ algorithm for a large Graph, Entries of $\mathbf{X}(t)$ versus Iterations $t$: Cauchy noise, $h(x)= \sqrt{\rho} \frac{2}{\pi} \tan^{-1}(\frac{\pi}{2} 0.01 x), f(x)= \tanh(5 x)$, $N=75$, $\xbar=134.31$, $\rho=5$ dB, $\gamma=0.1$.}\label{fig:TxRx_Cauchy_TanxArcTanx_LargeWSN_Noisy}
\end{center}
\end{minipage}
\end{figure}

\begin{figure}[tb]
\begin{minipage}{1\textwidth}
\centering
\begin{center}
\includegraphics[height=9.15cm,width=12cm]{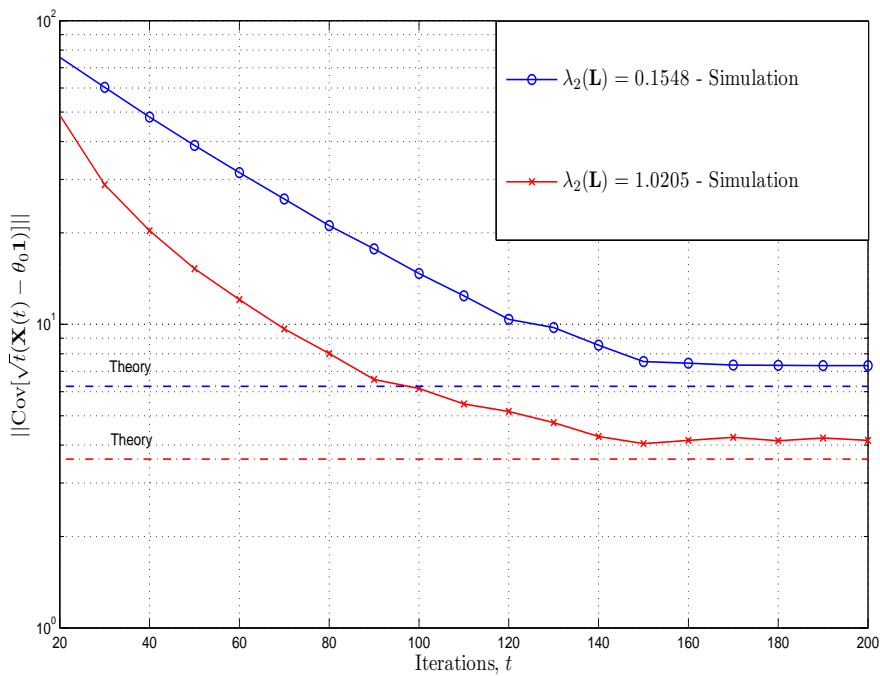} 
\caption{Difference in speed of convergence: sparsely versus densely connected Graphs, $||{\rm Cov} [ \sqrt{t}(\mathbf{X}(t)- \theta_0 \bf{1})]||$ versus Iterations $t$: Cauchy noise, $h(x)=x$, $f(x)=\frac{1.5 x} {1+ | 1.5 x|}$, $N=75$, $\theta_0=85.49$, $\xbar=84.31$, $\gamma=0.413$.}\label{fig:Diff_WSN_Same_Fx_Hx_Cauchy}
\end{center}
\end{minipage}
\end{figure}

\begin{figure}[tb]
\begin{minipage}{1\textwidth}
\centering
\begin{center}
\includegraphics[height=9.2cm,width=12cm]{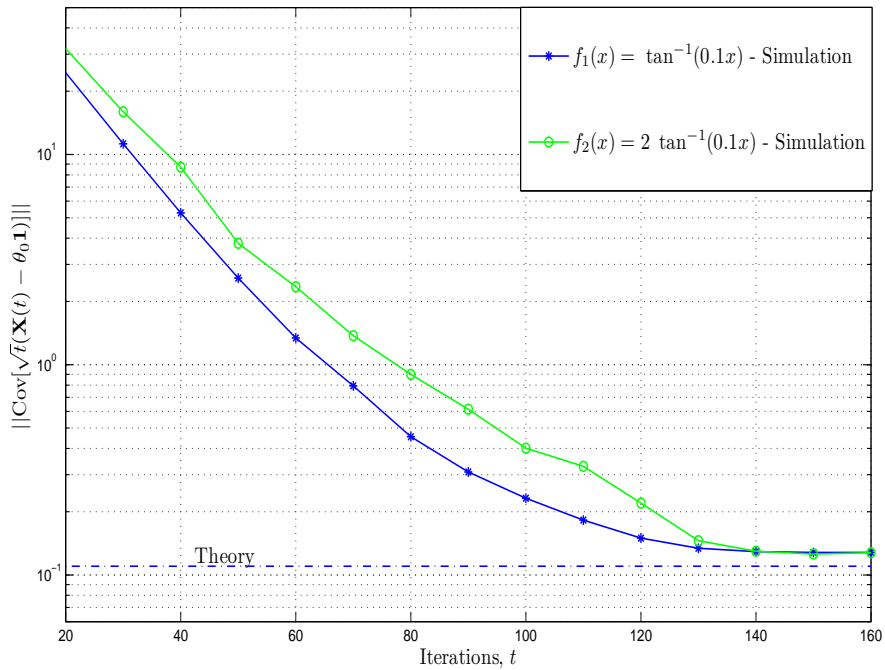} 
\caption{Scaling $f(x)$ does not change speed of convergence, $||{\rm Cov} [ \sqrt{t}(\mathbf{X}(t)- \theta_0 \bf{1})]||$ versus Iterations $t$: $h(x)=x$, $N=10$, $\theta_0=32.63$, $\xbar=34.31$, $\gamma=0.413$.}\label{fig:Rx_Tx_Scaling_Speed_SignxAbsx2}
\end{center}
\end{minipage}
\end{figure}

\begin{figure}[tb]
\begin{minipage}{1\textwidth}
\centering
\begin{center}
\includegraphics[height=9.2cm,width=12cm]{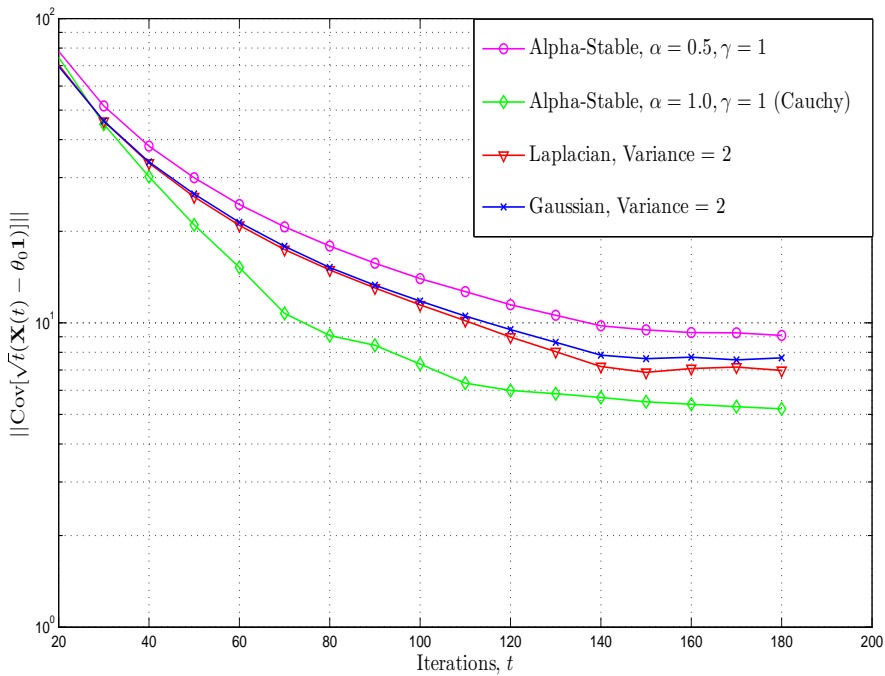} 
\caption{Robustness to various noise distributions, $||{\rm Cov} [ \sqrt{t}(\mathbf{X}(t)- \theta_0 \bf{1})]||$ versus Iterations $t$: $h(x)=x$, $f(x)= \tanh(2 x)$, $N=75$, $\theta_0=120.36$, $\xbar=124.31$}\label{fig:Diff_Noise_Distributions_Robustness3}
\end{center}
\end{minipage}
\end{figure}

\begin{figure}[tb]
\begin{minipage}{1\textwidth}
\centering
\begin{center}
\includegraphics[height=9.2cm,width=12cm]{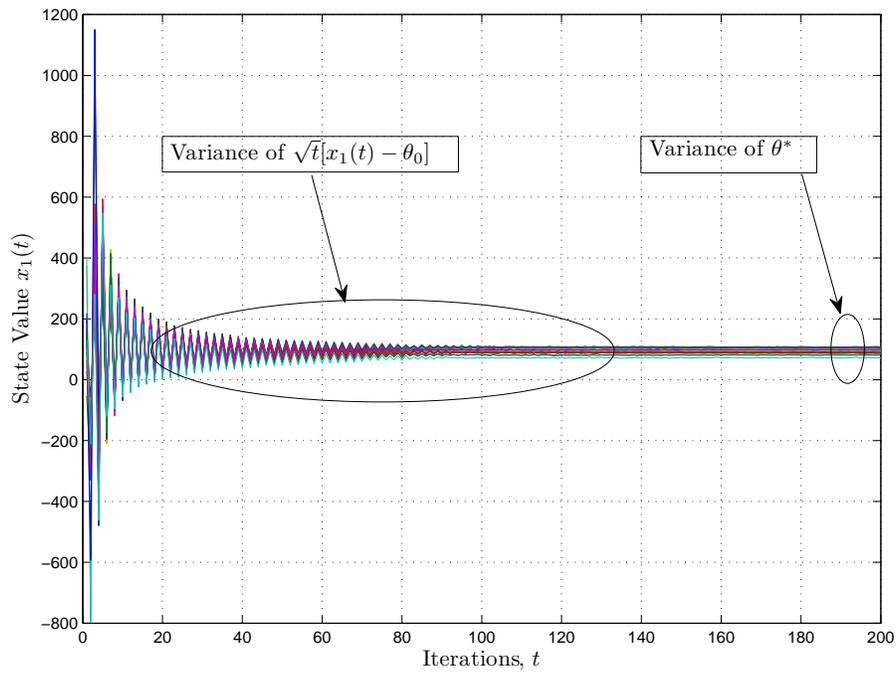}
\caption{Difference between Variance of $\cval$ and Asymptotic Variance, Entries of $\mathbf{X}(t)$ versus Iterations $t$: $h(x)=x$, $f(x)=3 \tan^{-1}(0.05 x)$, $N=75$, $\xbar=94.31$, $\gamma=0.413$}\label{fig:Difference_Between_VarTheta_Cov}
\end{center}
\end{minipage}
\end{figure}

Figures \ref{fig:TxRx_Cauchy_TanxArcTanx_SmallWSN_Noisy} - \ref{fig:Difference_Between_VarTheta_Cov} illustrate the performance of $\rnld$ algorithm in the presence of communication noise. As explained in the assumption \textbf{(A5)} in Section \ref{subsec:nld_with_noise_robust}, we chose the decreasing step sequence to be $\alpha(t)=1/(t+1), t \geq 0$, in all simulations. Here we assumed that $\rho=\max_{x} h^{2}(x)$ is the maximum power available at each sensor to transmit its state value. The receiver nonlinear function $f(x)$ is indicated in each case. Figure \ref{fig:TxRx_Cauchy_TanxArcTanx_SmallWSN_Noisy} shows that the nodes employing the $\rnld$ algorithm reach consensus for a small network with $N=10$ in about $100$ iterations and Figure \ref{fig:TxRx_Cauchy_TanxArcTanx_LargeWSN_Noisy} shows convergence for a large network with $N=75$ in about $40$ iterations.  

In Figures \ref{fig:Diff_WSN_Same_Fx_Hx_Cauchy}, \ref{fig:Rx_Tx_Scaling_Speed_SignxAbsx2} and \ref{fig:Diff_Noise_Distributions_Robustness3} we show the convergence speed performance of the proposed $\rnld$ algorithm by plotting the maximum eigenvalue of the covariance matrix of the vector process $\sqrt{t}(\mathbf{X}(t)- \theta_0 \bf{1})$ versus iterations $t$. These plots indicate how fast the process $\mathbf{X}(t)$ converges towards the limiting value $\theta_0 \onevect$. 

The speed of convergence for two graphs with different algebraic connectivity is illustrated in Figure \ref{fig:Diff_WSN_Same_Fx_Hx_Cauchy}. We see that the graph with smaller connectivity (smaller $\lambda_2(\La)$) converges slower than the one with large connectivity as dictated by  \eqref{eq:asymp_covariance_value_robust}. In Theorem \ref{asym_norm_nld_lemma_robust}, we also saw that scaling $f(x)$ does not change the asymptotic convergence speed. This is shown in Figure \ref{fig:Rx_Tx_Scaling_Speed_SignxAbsx2} where we see that when the iterations are large ($t > 140$), the speed of convergence of all the three functions are nearly the same. We depict the robustness of the $\rnld$ algorithm for various channel noise distributions in Figure \ref{fig:Diff_Noise_Distributions_Robustness3}. We observe that the performance is nearly the same for Gaussian and Laplacian distributions, whereas there is a significant gap between Cauchy and alpha-stable distributions considered in this simulation. The latter effect is due to the fact that, for a given $f(\cdot)$, the ratio $ \E_n [f^2(n)] / (\E_n[f^{'}(n)])^2 $ is significantly different for those two cases which justifies the performance gap. Finally, we illustrate the difference between the variance of $\cval$ and the asymptotic variance in Figure \ref{fig:Difference_Between_VarTheta_Cov}. Here we consider the evolution of the state value $x_1(t)$ of the first node for several consensus runs for the same initial conditions. Recall that in every consensus run the state value $x_1(t)$ converges to an instance of the limiting random variable $\cval$ and the variation among these several realizations is characterized by the variance of $\cval$. In contrast, how fast the state value $x_1(t)$ converges to the limiting value $\theta_{0}$ is characterized by the asymptotic variance of $\sqrt{t} [x_{1}(t) - \theta_{0}]$ as $t \rightarrow \infty$.

\section{Conclusions} \label{Sec:Conclusions:consensus_robust}
A distributed average consensus algorithm that converges in the presence of impulsive noise is considered. Every sensor also maps its state value through a bounded function before transmission to constrain the transmit power. It is shown that non-linearity at the receiver nodes makes the algorithm robust to a wide range of channel noise distributions including heavy-tailed channel noise. The proposed algorithm relaxes the requirement of finite moments on the communication noise and thus it is proved to be not only more general than the existing consensus algorithms but is practically viable for WSNs deployed in adverse conditions. It is proved using the theory of Markov processes that the sensors reach consensus asymptotically on a finite random variable whose expectation contains the desired sample average of the initial sensor measurements, and whose mean-squared error is bounded. The asymptotic convergence speed of the proposed algorithm is characterized by deriving the asymptotic covariance matrix using results from stochastic approximation theory. It is shown that the norm of the asymptotic covariance matrix is limited by the Fisher information of the noise distribution with respect to a location parameter.

\bibliographystyle{IEEEtran}
\bibliography{consensus}

\newpage

\end{document}